# Optimized Household Demand Management with Local Solar PV Generation

Seyyed Hamid Elyas, Hamidreza Sadeghian, Hayder O. Alwan, Zhifang Wang
Department of electric and computer Engineering
Virginia Commonwealth University
Richmond, VA, USA
Email: {elyassh, sadeghianh, alwanho, zfwang} @ vcu.edu

*Abstract* – Demand Side Management (DSM) strategies are often associated with the objectives of smoothing the load curve and reducing peak load. Although the future of demand side management is technically dependent on remote and automatic control of residential loads, the end-users play a significant role by shifting the use of appliances to the off-peak hours when they are exposed to Day-ahead market price. This paper proposes an optimum solution to the problem of scheduling of household demand side management in the presence of PV generation under a set of technical constraints such as dynamic electricity pricing and voltage deviation. The proposed solution is implemented based on the Clonal Selection Algorithm (CSA). This solution is evaluated through a set of scenarios and simulation results show that the proposed approach results in the reduction of electricity bills and the import of energy from the grid.

## NOMENCLATURE

Variables:

| | |
|---|---|
| $u_a$ | Binary status of appliance $a$; 0 = *off*, 1 = *on* |
| $r_a$ | Rated power in kW for appliance $a$ |
| $A$ | Total number of appliances |
| $A_{int}$ | Total number of interruptible appliances |
| $T$ | Total number of time slots, $T=48$ |
| $t$ | Index for time slots |
| $D_a$ | During of operation time for appliance $a$ |
| $[s_a, f_a]$ | Allowable operational time range for appliance $a$ |
| $t_a^{st}$ | Operation start time of appliance $a$ |
| $t_a^{en}$ | Operation end time of appliance $a$ |
| $t_a^{st_{new}}$ | Operation start time of appliance $a$ after DSM shifting |
| $t_a^{st_{old}}$ | Operation start time of appliance $a$ before shifting |
| $P_{pv}(t)$ | PV output in kW at time slot $t$ |
| $\pi_e(t)$ | Electricity price (¢/kWh) at time slot $t$ |
| $\pi_p$ | Penalty price (¢/kWh) |
| $\Delta T_a$ | Time shifting for appliance $a$ |
| $MD$ | Maximum demand of household power consumption |
| $B$ | Total number of buses |
| $b$ | Index for buses |
| $V(t)$ | Vector of bus voltage at time slot $t$ |
| $P(t)$ | Vector of active load power at t |
| $Q(t)$ | Vector reactive power at t |
| $f_{AC}(.)$ | AC power flow equation |
| $P_{loss}(t)$ | Active power loss in distribution feeder |
| $C_e$ | Total electricity cost (¢/day) |
| $C_p$ | Total penalty cost (¢/day) |

## I. INTRODUCTION

In recent years, the electricity networks started to change, with a wide range of distributed energy suppliers from wind turbine to photovoltaic systems. Along with distribution networks, modern communication infrastructures have also begun to be installed, in order to support and improve the reliability and efficiency of the power networks [1]. In this new electricity network, a large number of data sets are available for residential consumers to improve the energy consumption policies, by means of changing their habits in using household appliances [2]. In order to evaluate the effectiveness of an energy management system, electricity demand needs to be analyzed in a high-resolution fashion. This is required in order to identify which type of electricity activities can be modified without any weighty impact on the consumer's lifestyle and freedom [3]. During the last decade, various models have been proposed to define household demand side management strategies for improving the performance of the distribution networks. But the focuses and contributions of the models tends to be different. Reference [4], proposed a model that based on devices future usage, the consumer is able to optimally schedule home appliances activities for the next day and with the goal of minimizing the electricity bill. In [5], real-time monitoring system is presented as an effective way to improve the efficiency of different control methods in the energy management system. This model provides a great potential to control the activities of appliances especially the indoor temperature control devices. In reference [6], the design of a genetic algorithm based control method is presented in order to reduce the electricity bill but considering user freedom at home. According to the results presented in [6], this method can reduce the total electricity cost considering the real-time monitoring system and the electricity price. Authors in [7], define a method of welfare maximization in which the level of customers' social welfare, before and after changing the pricing policy from a flat pricing method to a real time one, remain unchanged. In reference [8], the authors propose an approach to model consumer demand at appliance level. Regarding the impact of renewable energy resources, many research works have been carried out in the literature [9]. A high penetration of distributed generation can lead to problems in low voltage distribution networks [10]. In [11], the authors proposed an active DSM model to address issues of integration of renewable energy resources in distribution networks, this paper considered uncertainties in load and power and proposed multi MGs power dispatch at smart distribution grids. Reference [12] provided a mathematical definition for electricity generation optimization in a typical residential load with different energy systems combined heat and a battery system.

The objective of this paper is to set up an optimization model for the offline household demand side management. The goal is to reschedule the energy consumption, taking into account the day-ahead dynamic electricity price [13] and the real production of the photovoltaic system. The proposed optimization-based model aims to reduce the total electricity bill but ensuring a comfortable user experience at home. This model can effectively minimize the energy consumption cost for day-ahead time horizon according to the forecasted electricity price. Compared to other related work in the literature, this work has following key contributions: 1) the proposed DSM model incorporates economic benefits of local solar PV generation along with negative impacts on voltage fluctuations and deviation in the distribution network. It should be noted here that most reference works ignore voltage problem in the presence of photovoltaic system. 2) This paper proposes a practical model for demand side management with a flexible penalty approach to account for the inconvenience caused by deviation from customer desired schedule. In other words, customer inconveniences caused by DSM schedule will be translated into additional compensation cost in the optimization objective function which is calculated based on some customized rate and intends to effectively discourage or reduce unnecessary load shifting or changes.

The rest of this paper is structured as follows. Section II gives the system model and mathematical description of demand side management problem. Section III provides and analyzes some numerical simulation results. Finally, section IV concludes the paper and discusses future works.

II. HOUSEHOLD DEMAND SIDE MANAGEMENT MODEL

In this section we will develop an optimization model for day-ahead household demand side management system with local solar PV generation. We first consider a very simple micro-grid network as shown in Fig. 1, which contains a single distribution feeder line supplying a small community of thirteen houses, each of which has a typical time-varying load profile generated by a time-series load modeling we built up based on some realistic residential customer load data obtained from the open-access database. A mid-size smart home is located at the end of the feeder equipped with a rooftop solar PV panel with a rated capacity of 6kW, which accounts for a 7.5% penetration rate given the total peak load of the system as 80 kW. In the future the optimization model developed based on this system may be expanded to include multiple feeders and scaled-up community size.

It is assumed that the smart home has a set of commonly used active appliances under a real-time pricing environment and the homeowner has access to day-ahead electricity rates and day-ahead forecast of PV generation and agrees to participate the DSM program to save his electricity bills with a controlled amount of convenience sacrifices. For the purpose of simplicity of analysis, thirty-one active appliances have been categorized into three groups based on their operational features: 1) Interruptible appliances: referring to electric devices allowed to switch on or off at any time during a day; 2) Uninterruptible appliances: referring to electric devices that need to operate until finish once started; and 3) Inflexible appliances: referring to appliances that will be active for entire simulation time (24 hour). Each appliance is modeled using four parameters $s_a, f_a, r_a$ and $D_a$, where $[s_a, f_a]$ defines the allowable operating time during which the appliance $a$ may be switched on, $r_a$ and $D_a$ denote the power rating and the total number of operating time slots as requested, respectively. Detailed operational data and constraints are shown in Table I.

Table I. Parameters of appliances used in the simulation

| Categorization of appliances | Appliance Index | $S_a \sim f_a$ | Original operation time slot | $D_a$ (30 min) | Power (kw) |
|---|---|---|---|---|---|
| Baseline appliances | 1 | 1 ~ 48 | 1 ~ 48 | 48 | 0.15 |
| | 2 | 1 ~ 48 | 1 ~ 48 | 48 | 1.60 |
| | 3 | 1 ~ 48 | 1 ~ 48 | 48 | 0.15 |
| Uninterruptible Appliances | 4 | 6 ~ 48 | 41 ~ 44 | 4 | 0.73 |
| | 5 | 6 ~ 30 | 18 ~ 20 | 3 | 0.73 |
| | 6 | 12 ~ 22 | 24 ~ 26 | 3 | 0.80 |
| | 7 | 12 ~ 22 | 27 ~ 28 | 2 | 0.80 |
| | 8 | 1~ 48 | 30~ 31 | 2 | 0.38 |
| | 9 | 1~ 48 | 15~ 18 | 4 | 0.38 |
| | 10 | 14~ 48 | 20~ 23 | 4 | 0.05 |
| Interruptible Appliances | 11 | 1 ~ 19 | 5 ~ 8 | 4 | 0.05 |
| | 12 | 1 ~ 48 | 39 ~ 42 | 4 | 1.26 |
| | 13 | 1 ~ 48 | 10 ~ 13 | 4 | 1.26 |
| | 14 | 1 ~ 48 | 33 ~ 36 | 4 | 0.70 |
| | 15 | 1 ~ 48 | 35 ~ 38 | 4 | 0.74 |
| | 16 | 1 ~ 48 | 22 ~ 25 | 4 | 0.64 |
| | 17 | 10 ~ 48 | 39 ~ 42 | 4 | 1.60 |
| | 18 | 10 ~ 48 | 32 ~ 35 | 4 | 1.90 |
| | 19 | 10 ~ 48 | 34 ~ 37 | 4 | 1.64 |
| | 20 | 10 ~ 48 | 36 ~ 39 | 4 | 1.50 |
| | 21 | 6 ~ 24 | 13 ~ 16 | 4 | 1.50 |
| | 22 | 10 ~ 48 | 35 ~ 39 | 4 | 1.50 |
| | 23 | 10 ~ 48 | 36 ~ 42 | 4 | 1.10 |
| | 24 | 10 ~40 | 15 ~18 | 4 | 2.00 |
| | 25 | 1 ~ 48 | 33 ~ 36 | 4 | 1.80 |
| | 26 | 1 ~ 44 | 18 ~ 21 | 4 | 0.25 |
| | 27 | 1 ~ 48 | 32 ~ 35 | 4 | 1.00 |
| | 28 | 1 ~ 48 | 12 ~ 15 | 4 | 1.20 |
| | 29 | 1 ~ 48 | 35 ~ 38 | 4 | 1.20 |
| | 30 | 1 ~ 48 | 17 ~ 20 | 4 | 1.20 |
| | 31 | 1 ~ 48 | 36 ~ 39 | 4 | 1.20 |

It is worth noting that in reality, each household may have a specific set of electric devices, with different power ratings and operational features. However, our proposed model assumes a certain homogeneity of household appliances to make a simpler configuration and more manageable implementation of DSM optimization. It should be mentioned here that in our model we tried to select a set of household appliances with higher power rate in order to show the impact of proposed DSM on single house model. In the future, we will aim at developing a model that includes multiple smart homes with more diversified combinations of appliances.

The PV generation curve used in this model is the real-power production data of a typical 6kW PV panel with some cloud effects, as shown in the Fig. 2. It should be mentioned here that the model has the capability of consuming its own produced energy locally if available and injecting the surplus power into the distribution network without any reward tariff. The operation cost of the PV panel system is negligible in this model. Note that the PV output power may drop to 50% of its rated power due to some weather conditions during a sunny day.

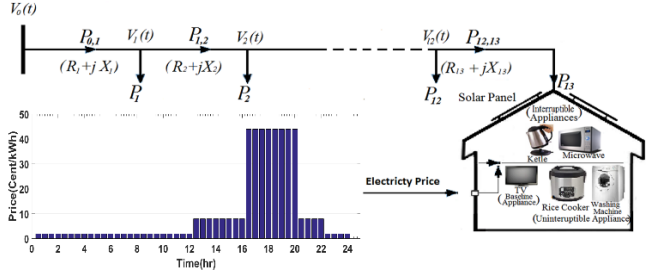

Fig.1. Hybrid system layout for a single house scenario

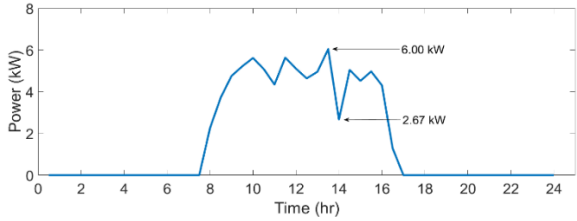

Fig. 2. Photovoltaic system generation

## III. MATHEMATICAL FORMULATION

The proposed household DSM model is aimed to minimize the electricity cost by scheduling the on/off status of domestic appliances over the operational periods, considering the dynamic electricity prices, locally available PV generation, and the penalty prices of appliance operation time-shifting which are included in order to manage the customer inconvenience caused by the proposed DSM program. Assume that the proposed demand side management program is scheduled day-ahead over a 24-h (30 min per slot) period. The decision variables are the operational status of appliances $u_a(t)$ over the next 24 hours and a typical schedule $[u_a(t)]_{A \times T}$ is a binary (0/1) matrix with following format:

$$[u_a(t)]_{A\times T} = [u_1^1, u_1^2, \ldots, u_1^T; u_2^1, u_2^2, \ldots u_2^T; \ldots; u_A^1, u_A^2, \ldots, u_A^T] \quad (1)$$

The objective function and constraints of the proposed DSM model can be presented as below:

$$\min_{u_a(t)} \mathcal{C}_e + \mathcal{C}_p \quad (2)$$

subject to:

$$\mathcal{C}_e = 0.5 \times \sum_{t=1}^{T}(P_{load}(t) + P_{loss}(t)) \times \pi_e(t) \quad (3)$$

$$\mathcal{C}_p = 0.5 \times \pi_p \sum_{a=1}^{A} \Delta T_a \, r_a \quad (4)$$

$$0 = f_{AC}\{P(t), Q(t), P_{pv}(t), V(t)\} \quad (5)$$

$$P_{loss}(t) = g_{AC}\{P(t), Q(t), P_{pv}(t), V(t)\} \quad (6)$$

$$P_{load}(t) = \max\left(\left(\sum_{a=1}^{A} r_a \times u_a(t) - P_{pv}(t)\right), 0\right) \quad (7)$$

$$0.95 \le |V_b(t)| \le 1.05 \quad \forall b \in \{1 \text{ to } B\} \quad (8)$$

$$\sum_{a=1}^{A} r_a \times u_a(t) \le MD \quad \forall a \in \{1 \text{ to } A\} \quad (9)$$

$$\sum_{t=1}^{T} u_a(t) = D_a \quad \forall a \in \{1 \text{ to } A\} \quad (10)$$

$$u_a(t) = 0 \quad \forall t < s_a \text{ or } \forall t > f_a \quad (11)$$

$$\Delta T_a = 1^T \cdot |t_a^{st_{new}} - t_a^{st_{old}}| \quad \forall a \in \{1 \text{ to } A\} \quad (12)$$

$$t_a^{st_{new}} = [t|u_a^{new}(t) = 1]_{1 \times D_a} \quad \forall a \in \{1 \text{ to } A_{int}\} \quad (13)$$

$$t_a^{st_{old}} = [t|u_a^{old}(t) = 1]_{1 \times D_a} \quad \forall a \in \{1 \text{ to } A_{int}\} \quad (14)$$

where (3)-(4) define the electricity cost and the penalty cost; (5)-(6) represent the distribution network constraints of AC power flow balance and real power loss. In our proposed model, surplus PV generation will be injected into the distribution without reward. Therefore, for house load at time $t$ we have assumed Eq. (7) to avoid negative electricity cost. In our proposed model, we have assumed that surplus PV generation will be injected into the distribution network without reward, so the total electricity cost within each time slot should be no less than zero. (8) gives the voltage constraints in the distribution network. Constraint (9) indicates the Maximum Demand (MD) that the aggregate appliance power of a household cannot exceed at any time. This upper is to prevent super-high power demand peak even during the hours when day-ahead electricity price is low because the utilities do not want to have "new" peak created by the DSM load-shifting or because the distribution feeders have capacity constraints. Constraint (10) and (11) indicate the total operation duration and the allowable turn-on time of an appliance. Constraints (12)-(14) specify the original and the new starting point, $t_a^{st_{old}}$ and $t_a^{st_{new}}$ respectively, to capture the duration of time-shifting for flexible appliances. It should be noted that an interruptible appliance, as illustrated in Fig.3, may have non-uniform time-shifting $\Delta T_a$ for all operation time slots, and in order to have a fair comparison, the time-shifting of each operation slot of an appliance should be counted in $\Delta T_a$.

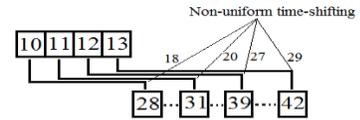

Fig. 3. Time-shifting in interruptible appliances

The aforementioned DSM optimization model will not simply move the operation times from peak-hours to off-peak time slots to save the electricity cost. A penalty cost included in the objective function (2) takes into account the customer willingness to change the appliances operation time. This means to discourage unnecessary load shifting that only causes small reduction in electricity cost ($\mathcal{C}_e$) than the increase from inconvenience-penalty cost ($\mathcal{C}_p$).

A heuristic algorithm called Clonal Selection Algorithm (CSA) is adopted to find the optimal operation schedule for the household appliances. The CSA is a powerful computational algorithm based on the biological immune system and the natural defense mechanism of human body. This method considers each candidate solution and its distance from global optimal solution, as an antibody and an antigen, respectively. The affinity of every single antibody is calculated via evaluation mechanism and then, they are sorted based on their affinity values. Finally, a new enhanced population is generated using immune operators [14]. In DSM problem, the optimal solution of this algorithm will determine the minimum total cost of electricity and time-shifting penalty, achieved with an optimal schedule $[u_a(t)]_{A\times T}$ which represents operation status of each appliance at each time slot.

## IV. NUMERICAL SIMULATIONS AND RESULTS

This section experiments the proposed household DSM approach over three scenarios in order to study its performance in terms of total household electricity and penalty cost, and the impact of PV generation on voltage deviation. In the base case scenario, the proposed DSM model is considered without penalty cost and without any PV installation. In the second scenario, the impact of different penalty costs on the DMS model is investigated without any PV generation. Finally, the last scenario is referred to the DSM optimization with consideration of the penalty cost and a local rooftop PV system of 6kW.

### A. DSM without penalty and without PV

The base case DSM in this section, with $\pi_p = 0$ and $P_{pv}(t) \equiv 0$, is modeled as a comparison benchmark for the other two scenarios. Fig.4 shows the original household consumption profile of the smart home without DSM scheduling with maximum peak demand 12.4 kW, which indicates two demand peaks around 7:00 to 9:00 in the morning and 17:00 to 20:00 in the evening. However, given the hourly electricity prices as shown in Fig.1, only the evening hours will experience high market prices therefore called "peak" hours. Without DSM program, the total electricity cost of this household is $18.68/day. If the proposed DSM program is implemented with an optimal appliance operation schedule, a new set of shifted consumption profile will be achieved with the total cost reduced to $5.99/day, as shown in Fig.5, where the red solid line represents original load profile. This accounts for about 68% saving in the electricity cost for the same amount of daily electric energy consumption. Fig.5 also indicates that most peak demands during high-price hours have been moved to off-peak periods except those of non-flexible appliances.

In the second scenario, we examine the impact of penalty price on the proposed DSM program. Fig. 6 (a)-(c) depict the optimal DSM consumption profiles at the penalty prices of $\pi_p = 5, 10$ and $20$ ¢/kWh respectively. Comparison of the DSM load profiles in Fig. 6(a)-(b) with that in Fig.5 shows that with penalty costs involved, most flexible loads will be been shifted to the medium-price hours at the boundary of peak hours because of the fact that larger time shifts lead to larger penalty cost given a constant penalty price of $\pi_p$ and may cancel some cost-saving benefits coming from the load shifting. As the penalty prices increases, more and more loads will not be moved. When the penalty price becomes $\pi_p = 20$ ¢/kWh, only a small amount of load has been shifted out of the peak hours. The main reason comes from the fact that if the cost-saving benefit, resulted from time shifting, becomes lower than corresponding penalty cost, then the suggested time shifting will be rejected.

It is worth noting that in order to avoid creating "new" peaks by the DSM load-shifting, an upper limit is put on the maximum peak demand in a household load profile at any hour during a day, with MD = 12.4 kW in our simulations. In DSM program with penalty, the MD constraint plays an important role in the scheduling of appliances. Fig. 6 (a) exhibits the impact of this constraint obviously. In this case the DSM program would reschedule some appliances with smaller time shifting hence with a lower penalty cost, if without the MD constraints. However, when the MD constraint in effect, the appliances have to be rescheduled with larger time shifting with higher penalty cost because the low-shifting hours cannot accommodate another appliance load without violating the MD constraint. For instance, with the penalty price $\pi_p = 5$ ¢/kWh, the load in medium price hour 16:00 has almost reached to the MD constraint. If we increase MD to 15 kW, more appliances will be rescheduled to hour 16:00; and with the penalty price $\pi_p = 10$ ¢/kWh we may shift more appliances from 17:00 to 19:30 if we set higher MD constraint.

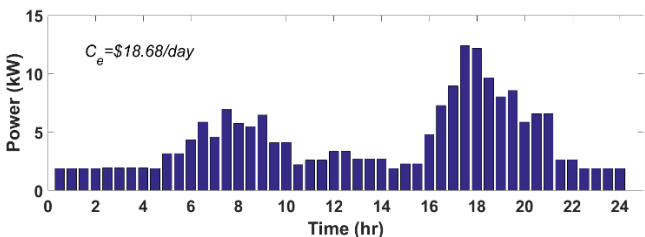
Fig. 4. Original household consumption profile without DSM

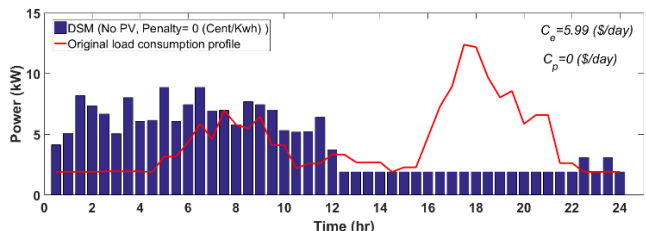
Fig.5. Consumption profile with DSM (No PV, No Penalty)

*B: DSM with Penalty and without PV*

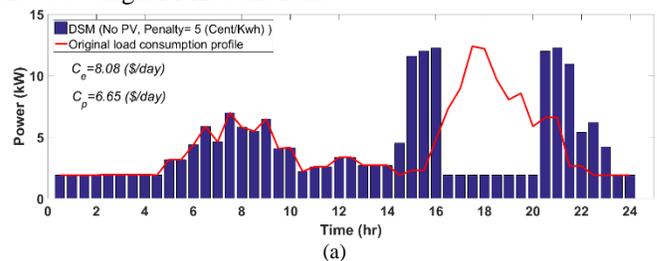
(a)

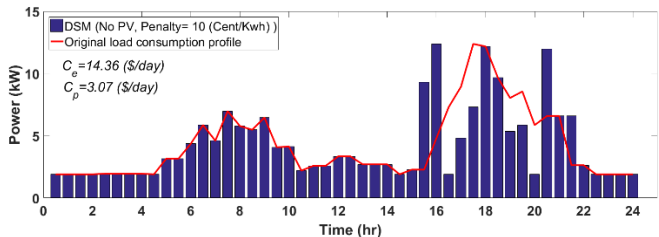
(b)

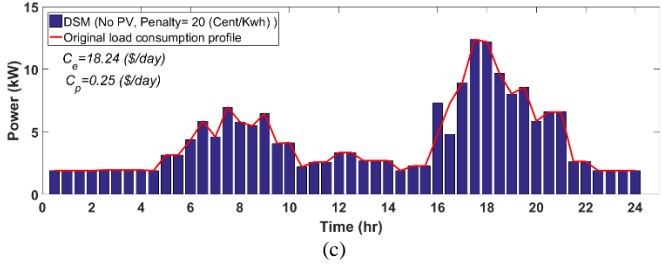

Fig.6. Consumption profiles with DSM at different penalty prices (No PV)

Fig. 7 presents the electricity cost savings, penalty cost, and time slot shifts of the flexible appliances participating in the DSM program with a penalty price $\pi_p = 5$¢/kWh. Fig. 7 indicates highly participation of interruptible appliance in the DSM program. It is worth noting that by increasing penalty factor, participation of uninterruptible appliances in the load shifting will be significantly discouraged due to their constraint on continuous operation after starting. For rescheduling uninterruptible appliances, DSM program needs to reschedule with more time shifting and most of time it impose larger penalty cost comparing with electricity cost saving. However, interruptible appliances with more flexibility to reschedule may actively participate in a DSM program with penalty factor because their operational feature allows to switch on or off at any time within the allowable operation time range. It should be mentioned that in this simulation scenario, some interruptible appliances originally operates at the lowest cost time slots, therefore the DSM program does not need to reschedule their operation times. Another fact is that the cost-saving benefit from an appliance not only depends on the electricity price decrease caused by its time shifting $\Delta T_a$ but also the appliance's rated power $r_a$. For example, Appliance 25 with 13 time slot shifting comparing has provided more cost-saving benefit for costumer with Appliance 27 with 12 time slot shifting, because of its higher rated power.

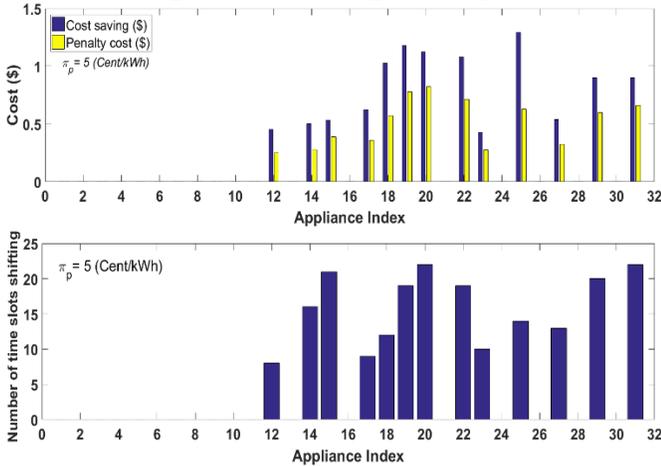

Fig.7. Cost saving and time shifting for DSM (No PV, With Penalty)

The simulation results indicate that, the daily electricity costs for three different penalty-based cases are $8.08, 14.36 and $18.24, and compared to the base-case consumption, the customer achieved 56.7%, 23.1% and 0.05% financial profit costs at the end of the day, respectively.

*C: DSM with PV generation*

This scenario is referred to the same house with the proposed DSM program and a 6 kW PV system. For this scenario, the PV generation curve illustrated in the Fig. 2 has been used. Making the best use of the installed PV can reduce the purchase of electricity in meeting the power demand. Therefore, PV generation utilization is an important consideration in assessing the scheduling performance, which can be defined as total PV generation used to meet the demand in comparison with total available PV generation. Fig.8 illustrates the DSM consumption profile in the presence of PV system with no penalty. By using DSM program, PV utilization efficiency is 99.98% for this case. Fig.8 shows that the available PV system is used at all the time slots except at 12:30, and 13:00 with a small amount of surplus PV generation. It is worthwhile to note that shifting appliance operation to this time slots exceeds the PV generation so, exceeded demand should be purchased with a price 8¢/kWh which it increases the total electricity bill. Considering the impact of PV system, the imported power from the grid is dramatically reduced and the customer has been allowed to operate more appliances from 9:00 to 16:00 (see Fig. 8). The obtained results show that the daily operation cost is reduced from $18.68/day to $4.48/day showing a reduction of 76 %.

Fig.9 shows the impact of different penalty costs on DSM program. As we can see, for $\pi_p = 5$ ¢/kWh the final consumption profile has a considerable difference from Fig.8. However, when the penalty cost becomes larger, the difference from the original consumption profile tends to decrease. It is worthwhile to note that due to PV generation, cost for operation at available PV time slots is zero, therefore by rescheduling appliances to this time periods we will have more cost savings. This fact can be seen by comparing Fig. 6 and Fig. 9 for $\pi_p = 5$ ¢/kWh . For DSM program with PV generation we can see more shifted appliances between 14:00 and 16:00 comparing with case without PV generation. For DSM with penalty 10 and 20 (Cent/kWh) we can see surplus PV generation generally between 10:00 and 14:30. For this cases PV utilization efficiency decrease to 72.41% and 68.63%, respectively. It means that higher penalty factors decrease costumer opportunity to use free and clean PV generation power.

Fig.10 illustrates the impact of installed PV system on the voltage at the smart home. Fig. 10 shows that considering larger penalty factor for appliance operation time shifting restricts the total number of time shifting. As a result, operation times for shifted appliance will concentrate on nearby time slots which it results in to appear two other peak loads and voltage drops in consumption load profile in comparison with original load profile with one peak load in peak hours. In addition, we can see an increase in voltage level along the feeder around noon which delivers a smoother voltage profile and improves voltage quality. In this time period we have PV generation, therefore, it decreases total net injected power from electricity network for supplying operation of appliances.

Table II illustrates the summary of obtained results for all three scenarios. As it mentioned, penalty factor models customer willingness to participate in the DSM program. We can see that by increasing penalty factor, household electricity costs increase to reach the original household cost meaning that with higher penalty factor household load profile does not experience a big difference with original load profile.

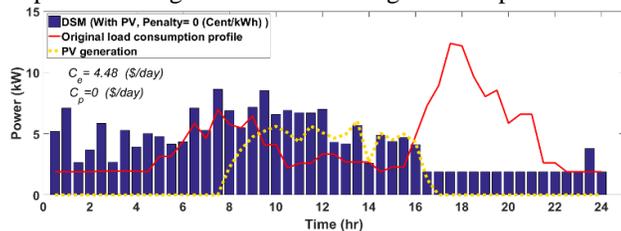

Fig. 8. Consumption profile with DSM (with PV, No Penalty)

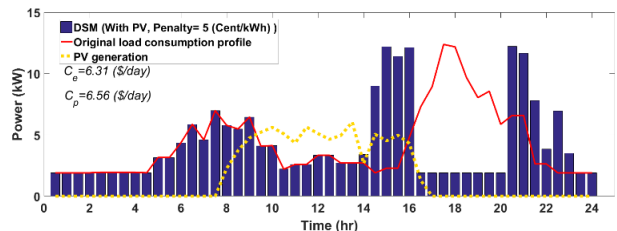

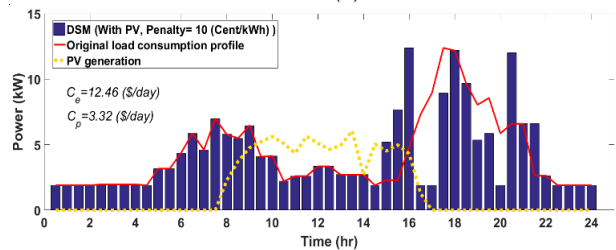

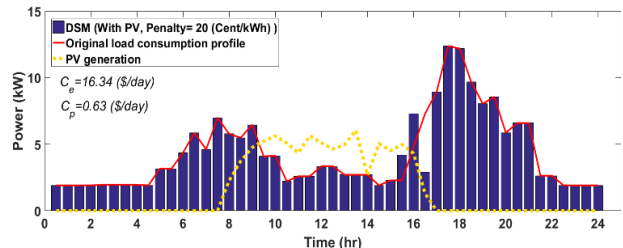

Fig.9. Consumption profiles for different penalty costs with DSM and PV

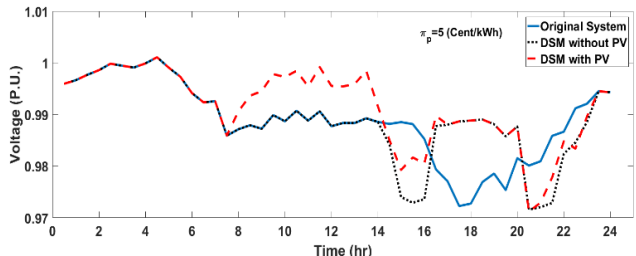

Fig. 10. Voltage profile at smart home with DSM

Table II. Comparison of Costs for the three Scenarios

| Scenario | $\pi_p$ [¢/kW] | Total Cost [$/day] | $c_e$ [$/day] | $c_p$ [$/day] |
|---|---|---|---|---|
| DSM (No PV, No penalty cost) | 0 | 5.99 | 5.99 | 0 |
| DSM (No PV, with penalty cost) | 5 | 14.73 | 8.08 | 6.65 |
|  | 10 | 17.43 | 14.36 | 3.07 |
|  | 20 | 18.49 | 18.24 | 0.25 |
| DSM (with PV, with penalty cost) | 0 | 4.48 | 4.48 | 0 |
|  | 5 | 12.87 | 6.31 | 6.56 |
|  | 10 | 15.78 | 12.46 | 3.32 |
|  | 20 | 16.97 | 16.34 | 0.63 |

## V. CONCLUSION AND FUTURE WORKS

In this paper, an approach has been presented for demand side management in smart residential homes. We first introduce the architecture of DSM in residential areas and then provide a practical approach for day ahead optimal scheduling of household appliances based on dynamic pricing and renewable generation forecasting. Our numerical results show that an effective demand side management provides benefits not only to the end users but also to the utilities by reducing the peak load demand and overall cost without violating the voltage deviation constraints. Our proposed approach can be used in demand side management systems to help household owners to automatically create optimal load operation schedules based on comfort settings and in the presence of dynamic electricity pricing and PV system. In this paper, we just consider a single photovoltaic system connected to the end of the feeder. In the future, further experiments and verification will be performed on realistic micro-grids and multiple households.